\documentclass[aps,prl,twocolumn,superscriptaddress,showpacs]{revtex4-1}

\usepackage{amsmath}
\usepackage{graphicx}
\usepackage{subfigure}
\usepackage[ansinew]{inputenc}

\newcommand{\ud}{\mathrm{d}}

\newcommand{\nn}{\nonumber}
\newcommand{\ob}[1]{\overline{#1}}

\renewcommand{\Re}[1]{\mathrm{Re}\left\{#1\right\}}
\renewcommand{\Im}[1]{\mathrm{Im}\left\{#1\right\}}

\newcommand{\mbR}{\mathbf{R}}
\newcommand{\mbe}{\mathbf{e}}
\newcommand{\mbE}{\mathbf{E}}

\newcommand{\mbk}{\mathbf{k}}

\newcommand{\mbm}{\mbox{\boldmath$\mu$}}

\newcommand{\mbG}{\mathbf{G}}

\newcommand{\OR}{\Omega_{\mathrm{R}}}
\newcommand{\De}{\Delta}

\newcommand{\Ht}{\mathcal{H}} 
\newcommand{\Ha}{\mathcal{H}_\mathrm{A}} 
\newcommand{\Hf}{\mathcal{H}_\mathrm{F}} 
\newcommand{\Hi}{\mathcal{H}_\mathrm{I}} 
\newcommand{\ha}{\hat{a}} 
\newcommand{\had}{\hat{a}^{\dagger}} 
\newcommand{\hS}{\hat{S}} 
\newcommand{\hbE}{\hat{\mathbf{E}}}

\newcommand{\hF}{\hat{F}}

\newcommand{\hbm}{\hat{\mbox{\boldmath$\mu$}}}

\newcommand{\bk}[3]{\left<#1\left|#2\right|#3\right>} 
\newcommand{\bt}[1]{\left<#1\right>} 
\newcommand{\kb}[2]{\left|#1\right>\left<#2\right|} 
\newcommand{\ket}[1]{\left|#1\right>} 

\renewcommand{\section}[1]{{\par\it #1.---}}
\begin{document}

\title{Cooperative Fluorescence from a Strongly Driven Dilute Cloud of Atoms}


\author{J.~R.~\surname{Ott}}
\altaffiliation[Present address: ]{Department of Theoretical Physics, University of Geneva, CH-1211 Geneva, Switzerland}\email{johan.ott@unige.ch}
\affiliation{Department of Photonics Engineering, Technical University of Denmark, DK-2800 Kgs. Lyngby, Denmark}
\affiliation{Universit{\'e} de Nice Sophia Antipolis, CNRS, Institut Non-Lin{\'e}aire de Nice, UMR 7335, F-06560 Valbonne, France}

\author{M.~\surname{Wubs}}
\affiliation{Department of Photonics Engineering, Technical University of Denmark, DK-2800 Kgs. Lyngby, Denmark}

\author{P.~\surname{Lodahl}}
\affiliation{Niels Bohr Institute, University of Copenhagen, Blegdamsvej 17, DK-2100 Copenhagen, Denmark}

\author{N.~A.~\surname{Mortensen}}\email{asger@mailaps.org}
\affiliation{Department of Photonics Engineering, Technical University of Denmark, DK-2800 Kgs. Lyngby, Denmark}

\author{R.~\surname{Kaiser}}
\affiliation{Universit{\'e} de Nice Sophia Antipolis, CNRS, Institut Non-Lin{\'e}aire de Nice, UMR 7335, F-06560 Valbonne, France}


\date{\today}

\begin{abstract}
\noindent We investigate cooperative fluorescence in a dilute cloud of strongly driven two-level emitters. Starting from the Heisenberg equations of motion, we compute the first-order scattering corrections to the saturation of the excited-state population and to the resonance-fluorescence spectrum, which both require going beyond the state-of-the-art linear-optics approach to describe collective phenomena. A dipole blockade is observed due to long range dipole-dipole coupling that vanishes at stronger driving fields. Furthermore, we compute the inelastic component of the light scattered by a cloud of many atoms and find that the Mollow triplet is affected by cooperativity. In a lobe around the forward direction, the inelastic Mollow triplet develops a spectral asymmetry, observable under experimental conditions.
\end{abstract}


\maketitle

Experimental progress in controlling light-matter interaction, e.g., in cold atomic clouds and solid state devices, has in recent years given rise to several proposals and demonstrations of using collections of atoms for quantum-information processing~\cite{ladd_nature10a, saffman_rmp10a}. When strongly driven, a single two-level emitter exhibits a spectral triplet, the so-called Mollow triplet~\cite{mollow_pr69a}, which, e.g., has been used for generation of heralded single photons and entangled photons from solid state quantum dots~\cite{ulhaq_nphot12a}. In collections of many emitters, cooperative phenomena induced by interatomic dipole-dipole interaction have been predicted for weak or no driving leading to cooperative decay rates~\cite{dicke_pr54a, courteille_epjd10a} and modified Lamb shifts~\cite{friedberg_prep73a,Friedberg2010} that have been observed experimentally~\cite{feld_prl73,rohlsberger_science10a,keaveney_prl12a}. While the combination of large collections of emitters and strong light-matter interaction is surely realizable in the laboratory, theory is faced by a challenge; this setting is difficult to tackle theoretically due to the complex nature of the non-linear many-body problem. 

State-of-the-art quantum-electrodynamics theory of driven atomic clouds typically consider the decay of initially inverted systems~\cite{dicke_pr54a,feld_prl73}, single-photon excitations in the many-atom case~\cite{scully_prl06a, svidzinsky_prl08a, svidzinsky_pra10a, scully_prl09a, courteille_epjd10a, bienaime_prl10a, bachelard_pra11a, bienaime_fsc12a}, few strongly driven atoms~\cite{agarwal_pra77a,savels_prl07a,das_prl08a}, or interference of light emitted by strongly driven non-interacting atoms~\cite{Jin_2011a}. 

In this Letter, we report how interatomic interactions influence the saturation of the excited-state population and the cooperative fluorescence spectrum of a strongly driven cloud of two-level atoms. Surprisingly, our results show that even when the dipole-dipole interaction between any pair of atoms is weak, such as in dilute clouds, the collective interatomic coupling is important for the nonlinear response.

\begin{figure}[t!]
	\centering
		\includegraphics[width=0.45\textwidth]{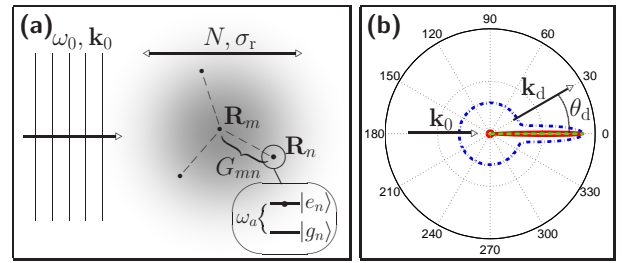}
	\caption{(Color online) (a) Sketch of the setting. A homogeneous laser field of frequency $\omega_0$ is incident in the direction of wavevector $\mbk_0$ onto a spherical Gaussian-distributed cloud of size $\sigma_{\rm r}$ containing $N$ identical atoms. Atoms at positions $\mbR_{m}$ and $\mbR_{n}$ are coupled via the radiation field through $G_{mn}$.  (b) Forward-directed elastic part of the fluorescence spectrum divided by $N^2$, calculated using Eq.~\eqref{eqn:Spec} with $N=3$~(blue dash-dotted line), $N=30$~(red solid line), and $N=30.000$~(green dashed line) and fixed $b_0=0.1$ pumped with $\De/\Gamma=-2.5$ and $\OR/\Gamma=5$. The emitted light is detected in the far field in the direction of the wavevector of the scattered light $\mbk_{\rm d}$ at an angle $\theta_{\rm d}$.}
	\label{fig:cloud}
\end{figure}

\section{Model} 
We consider $N$ identical two-level atoms, where the $m^{\rm th}$ atom at position $\mbR_m$ has a ground state $\ket{g_m}$, and an excited state $\ket{e_m}$, separated by the transition energy $\hbar \omega_a$. The atoms are driven by a plane-wave laser with wavevector $\mbk_0$, frequency $\omega_0$, and amplitude $\mbE_0$, see Fig.~\ref{fig:cloud}(a). The full Hamiltonian has the form $\Ht=\Ha+\Hf+\Hi$. Here, $\Ha$ is the free atomic Hamiltonian $\Ha=\sum_m\hbar\omega_{\rm a}\hS_m^{z}(t)$, where $\hS_m^{z}(t)=\frac{1}{2}(\kb{e_m}{e_m}-\kb{g_m}{g_m})$ is the population-inversion operator of the $m^{\rm th}$ atom. The free-field Hamiltonian is $\Hf=\sum_\lambda\hbar\omega_\lambda\had_\lambda(t)\ha_\lambda(t)$, where $\ha_\lambda(t)$ is the bosonic annihilation operator of the photonic mode $\lambda$ with frequency $\omega_\lambda$. The electric-dipole Hamiltonian $\Hi=-\sum_m \hbm_m(t)\cdot\hbE(\mbR_m,t)$ describes the light-matter interaction. Here $\hbm_m(t)=\mbm_m^*\hS_m^{+}(t)e^{i\omega_0t-i\mbk_0\cdot\mbR_m} +H.c.$ is the dipole operator, $\hS_m^{+}(t)=\kb{e_m}{g_m}\exp(i\mbk_0\cdot\mbR_m -i\omega_0t)$ the raising operator of the $m^{\rm th}$ atom rotating in the frame of the incident field, $\hS_m^{-}(t)=[\hS_m^{+}(t)]^{\dag}$ the corresponding lowering operator, and $\mbm_m=\bk{g_m}{\hbm_m}{e_m}$ is the dipole moment. Finally, $\hbE(\mbR_m,t)=i\sum_\lambda g_\lambda \mbe_\lambda e^{i\mbk_\lambda\cdot\mbR_m} \ha_\lambda(t)+H.c.$ is the electric-field operator with $\mbe_\lambda$ the polarization vector of mode $\lambda$ and $g_\lambda=\sqrt{\frac{\hbar\omega_\lambda}{2\epsilon_0}}$ where $\epsilon_0$ is the vacuum permittivity.

\section{Dynamics}
We work in the Heisenberg picture and after the Born--Markov approximation arrive at the equations of motion for the operators of the $m^{\rm th}$ atom~\cite{lehmberg_pra70a}
\begin{subequations}
\begin{align}
\frac{\ud }{\ud t} \hS^{+}_m&=-(\Gamma/2+i\Delta)\hS^{+}_m+i\OR\hS^{z}_m\nn\\
&\qquad\qquad+2i\sum_{n\neq m}G_{mn}^*\hS_n^{+}\hS^{z}_m+\hF_m^{+},\\
\frac{\ud }{\ud t} \hS^{z}_m&=-\Gamma\left(\hS^{z}_m+\frac{1}{2}\right)+\frac{i\OR}{2}\left(\hS^{+}_m-\hS^{-}_m\right)\nn\\
&\quad+i\sum_{n\neq m}\left(G_{mn}\hS_m^{+}\hS^{-}_n-H.c.\right)+\hF_m^{z}.
\end{align}\label{eqs:S}
\end{subequations}
Here, $\OR=|\mbm_m\cdot \mbE_0|/\hbar$ is the Rabi frequency, $\Gamma=4\mu^2\omega_{\rm a}^3/(3\hbar c^3)$ the spontaneous-decay rate, $\mu=|\mbm_m|$ is the magnitude of the dipole moment that is equal for all atoms, $c$ the speed of light in vacuum, $\Delta=\omega_0-\omega_{\rm a}-\eta$ the detuning between the driving field and the atomic resonance, and $\eta$ the Lamb shift. Eqs.~\eqref{eqs:S} are derived in the rotating-wave approximation for the atomic operators. However the counter-rotating terms in the interaction Hamiltonian are maintained in order to make a useful connection with classical optics~\cite{wubs_pra04b,lehmberg_pra70a,svidzinsky_pra10a}, since then the  dipole-dipole coupling terms $G_{mn}$ are related to the classical Green tensor $\mbG$~\cite{de_vries_rmp98a} by
\begin{align}
G_{mn}&=-\frac{\mu_0\omega_{\rm a}^2}{\hbar}\mbm_m^*\cdot\mbG(\mbR_{mn},\omega_{\rm a})\cdot\mbm_n e^{-i\mbk_0\cdot\mbR_{mn}}\nn\\
&=\frac{\Gamma}{2} \frac{e^{ik_{\rm a}R_{mn}}}{k_{\rm a}R_{mn}}e^{-i\mbk_0\cdot\mbR_{mn}},\label{eqn:Gmn}
\end{align}
where $\mbR_{mn}=\mbR_m-\mbR_n$, $R_{mn}=|\mbR_{mn}|$, $k_{\rm a}=\omega_a/c$, and we use the scalar model for $\mbG$ that is justified for dipole-dipole coupling in the case of dilute clouds~\cite{Friedberg2010}. Finally, the terms $\hF^{+}$ and $\hF^{z}$ in Eqs.~\eqref{eqs:S} are the Langevin operators, which are given by normal-ordered combinations of products of atomic and field operators~\cite{CCohen-Tannoudji_1989,footnote:EPAPS}.

Eqs.~\eqref{eqs:S} describe the quantum nonlinear dynamics of a cloud of atoms driven by a plane wave of light. For a single atom, these equations reduce to the well-known optical Bloch--Langevin equations~\cite{CCohen-Tannoudji_1989}. Another simple limit of Eqs.~\eqref{eqs:S} that does not suffice for the present work is the linear-optics limit. The linear dynamics of Refs.~\cite{scully_prl06a, svidzinsky_prl08a, svidzinsky_pra10a, scully_prl09a, courteille_epjd10a, bienaime_prl10a, bachelard_pra11a, bienaime_fsc12a} is obtained from Eqs.~\eqref{eqs:S} by the usual approximation $\hS_m^{(z)} = -\frac{1}{2}$, valid for weak driving ($\Omega_{\rm R}/\Gamma\ll 1$) that maintains the atoms  mainly in their ground states. Furthermore, by letting $\OR=0$, Eqs.~\eqref{eqs:S} also describe the dynamics of initially inverted systems leading to superfluorescence as investigated in Refs.~\cite{dicke_pr54a,feld_prl73}. Here we focus on nonlinear quantum cooperative effects due to strong driving.

\section{Approximate solutions and validity for dilute clouds}
Solving Eqs.~\eqref{eqs:S} for the expectation values scales as $4^{N}$ so that exact numerical computations for clouds having say $N \gg 100$ are beyond reach. In the following we focus on dilute clouds and aim for accurate rather than exact dynamics. This allows the simplifying approximation that the dipole-dipole coupling between any two atoms is small, {\em i.e.}, $G_{mn}$ is treated as a perturbation to first order. This approximation greatly simplifies the problem and allows for analytic expressions for the expectation values of single-time operators. Furthermore, for two-time correlations the Langevin terms contribute negligibly and two-time dynamics can thus be reduced by the quantum regression theorem to single-time dynamics. Some details of the method and calculations can be found in Ref.~\cite{footnote:EPAPS}. The approximate solutions are  valid for small optical thickness, $b_0=3N/(k_0\sigma_{\rm r})^2\ll 1$. This is a more severe restriction than on linear theories~\cite{scully_prl06a, svidzinsky_prl08a, svidzinsky_pra10a, scully_prl09a, courteille_epjd10a, bienaime_prl10a, bachelard_pra11a, bienaime_fsc12a} that are valid for small off-resonance optical thickness $b_{\De} = b_0/(1 + 4\De^2/\Gamma^2)\ll 1$~\cite{bachelard_pra11a,bienaime_fsc12a}. This difference in range of validity can be understood by the fact that, contrary to the linear theories, our approach takes all frequencies into account and thus there will always be some part of the spectrum which is in resonance with the atomic transition energy.

\section{Steady-state population}
Let us first calculate the steady-state population of the $m^{\rm th}$ atom, $n_m=\langle\hS_m^{z}\rangle+1/2$. Let $n_m = n_m^{(0)}+n_m^{(1)}$, where $n_m^{(0)}=s/[2(1+s)]$ is the usual single-atom population, expressed in terms of the  saturation parameter $s=\OR^2/[2(\Gamma^2/4+\Delta^2)]$, and $n_m^{(1)}$ is the first-order correction due to the dipole-dipole interactions. Solving Eqs.~\eqref{eqs:S} as a matrix equation to first order in $G_{mn}$, we arrive at~\cite{footnote:EPAPS}
\begin{align}
\label{eqs:nm}
n_m = \frac{s}{2(1+s)}-\frac{(\Im{G_m}\Gamma/2+\Re{G_m}\De)s}{(\Gamma^2/4+\De^2)(1+s)^3},
\end{align}
where $G_m=\sum_{n\neq m}G_{mn}$. There is an interesting connection between $G_m$, and the cooperative decay rate, $\Gamma_N$, and Lamb shift, $\eta_N$: By averaging over atomic positions (denoted by an overbar) and considering a spherical Gaussian-distributed atomic cloud of root-mean-square size $\sigma_{\rm r}$ (corresponding to atoms in a harmonic potential), we obtain for $k_0\sigma_{\rm r}\gg1$ that $\ob{\Im{G_m}}=\Gamma (N-1)/[2(2k_0\sigma_{\rm r})^2]$ and  $\ob{\Re{G_m}}=\Gamma(N-1)/[2\sqrt{\pi}(2k_0\sigma_{\rm r})^3]$~\cite{footnote:EPAPS}. These are respectively $\Gamma_N$ and $\eta_N$ e.g. found from single-photon scattering~\cite{scully_prl09a,courteille_epjd10a} and $\eta_N$ also from the scattering correction to the expectation value of the Hamiltonian~\cite{friedberg_prep73a,Friedberg2010}.

\begin{figure}[t!]
	\centering
		\includegraphics[width=0.45\textwidth]{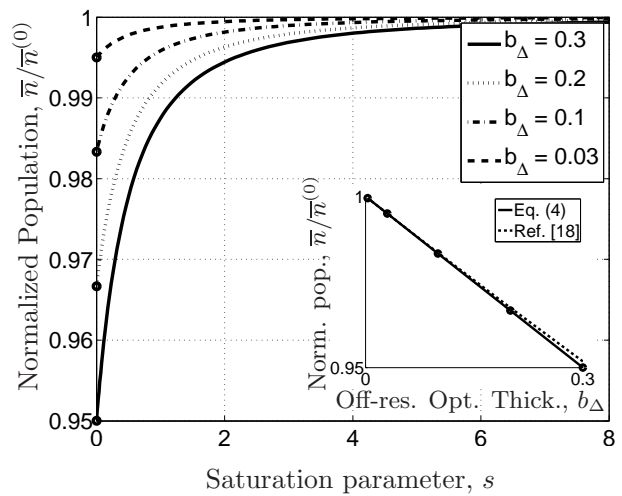}
	\caption{Normalized population based on Eq.~(\ref{dipole_blockade}) as a function of saturation parameter, $s$, for various values of the off-resonance optical thicknesses $b_{\Delta}$. The circles correspond to the small saturation parameters used in the inset. Inset: Comparison of $\ob{n}/\ob{n}^{(0)}$ based on Eq.~(\ref{dipole_blockade}) for weak driving ($s=2\times10^{-12}$) with the linearized single-photon multiple-scattering theory of Ref.~\cite{bienaime_fsc12a}, as a function of off-resonant optical thickness $b_{\De}$.}
	\label{fig:norm_pop_w_b0_inset}
\end{figure}

For $k_0\sigma_{\rm r}\gg 1$, \emph{i.e.}, large clouds, $\Gamma_N/\eta_N=\sqrt{\pi}k_0\sigma_{\rm r}$ such that $\eta_N$ is negligible and we obtain as a main result that the ensemble-averaged mean excited-state population, $\ob{n}=\frac{1}{N}\ob{\sum_mn_m}=\ob{n}^{(0)}+\ob{n}^{(1)}$, is given by
\begin{align}\label{dipole_blockade}
\ob{n} &\approx \frac{s}{2(1+s)}-\frac{b_{\Delta}s}{12(1+s)^3},
\end{align}
expressed in terms of the off-resonant optical thickness $b_{\De}$. Eq.~\eqref{dipole_blockade} shows that for a Gaussian cloud $\ob{n}^{(1)}$ is always negative, in other words the dipole-dipole interactions decrease the steady-state population. This can be interpreted as a cooperative dipole blockade implying that the presence of other atoms in the cloud leads to a less efficient excitation of the emitters. The effect is illustrated in Fig.~\ref{fig:norm_pop_w_b0_inset}, showing that the steady-state excited-state population decreases with increasing optical thicknesses. The nonlinear monotonous increase of the population with $s$ illustrates that stronger driving makes dipole-dipole interactions less important relative to the interaction with the driving field, and for $\Omega_{\rm R}\gg \Gamma$ we recover the steady-state population of noninteracting atoms. This agrees with and generalizes theoretical observations for two and three atoms~\cite{das_prl08a,agarwal_pra77a}.

To further corroborate our results, we show in the inset of Fig.~\ref{fig:norm_pop_w_b0_inset} that Eq.~(\ref{dipole_blockade}), agrees with existing single-photon multiple-scattering theory~\cite{bienaime_fsc12a} in the limit of weak scattering. In more detail, to lowest order in $s$ and $b_0$ Eq.~(\ref{dipole_blockade}) becomes $\ob{n}/\ob{n}^{(0)} = 1-\frac{b_\De}{6}$, in agreement with Ref.~\cite{bienaime_fsc12a}. Thus, Eq.~(\ref{dipole_blockade}) unifies both the known dipole-blockade effect for weak driving and the novel inclusion of saturation effects of the dipole-blockade for strong driving.

It is interesting to note that, while the large Gaussian cloud considered here always results in a blockade effect, \emph{i.e.} $\ob{n}^{(1)}<0$, an \emph{enhanced} population due to cooperative coupling could be obtained by either of two ways: $i$) If $\eta_N\De$ dominates over $\Gamma_N\Gamma/2$ a transition from negative to positive first-order correction $n^{(1)}$ is obtained by varying the detuning. $ii$) If $\Gamma_N\Gamma/2$ is negative. The case $i$) is, e.g., obtained for a Gaussian cloud when $|\De|/\sqrt{\pi}\Gamma>k_0\sigma$ which for a cloud-size of $k_0\sigma_{\rm r}\sim 50$ would need a detuning of $\De/\Gamma\sim-100$. The case $ii)$ could be obtained by controlling the atomic positions (e.g., with an optical lattice) since the real and imaginary parts of $G_{mn}$ both oscillate around zero as a function of interatomic distance and thus careful positioning could give a negative $\Gamma_N=\ob{\Im{G_m}}$.

\section{Fluorescence spectrum}
Next as our main investigation we study the effect of the dipole-dipole interactions on the steady-state resonance-fluorescence spectrum of the atomic cloud. By assuming that non-scattered light is filtered out, we can write the far-field spectrum at detection angle $\theta_{\rm d}$ as 
\begin{multline}
S(\theta_{\rm d},\omega)/S_0= \\
\sum_{m,n}\Re{\lim_{t\rightarrow \infty}\int_{0}^{\infty}\ud \tau \bt{\hS_m^{+}(t+\tau)\hS_n^{-}(t)}e^{i\delta\omega\tau -i\delta\mbk\cdot\mbR_{mn}}},\label{eqn:Spec}
\end{multline}
with $S_0=k_0^4\mu^2/(12\pi^2\epsilon_0^2r^2)$ where $r$ is the distance from the center of the cloud to the detector, $\delta\omega=\omega-\omega_0$, $\delta\mbk=\mbk_{\rm d}-\mbk_0$, and $\mbk_{\rm d}$ is the wave-vector of photons in the detection direction. The spectrum consists of two parts. The terms with $m=n$ in Eq.~\eqref{eqn:Spec} concern photons emitted from the $N$ individual atoms. The $m\neq n$ terms correspond to interference between photons emitted from different atoms. We emphasize that both intensity and interference parts have collective features, as the excitation of each atom is self-consistently obtained by considering the drive by the total field, {\em i.e.} the incident field plus the field scattered by all the other atoms. We evaluate Eq.~\eqref{eqn:Spec} using the quantum-regression theorem and split the result into the elastic spectrum $S_{\rm el}$ as well as the inelastic spectrum $S_{\rm in}$, which we discuss separately below.

\section{Elastic spectrum}
Based on Eqs.~\eqref{eqs:S} and~\eqref{eqn:Spec}, we calculate the ensemble-averaged angular-emission pattern of the elastic spectrum, $\ob{S_{\rm el}(\theta_{\rm d})}$. It consists of an isotropic part, corresponding to the intensity emission, and a strongly forwardly directed lobe, due to the interference part of the spectrum~\cite{footnote:EPAPS}. Close to the forward direction, $\ob{S_{\rm el}(\theta_{\rm d})}$ scales as $f^2(\theta_{\rm d})=\exp\{-2[k_0\sigma_{\rm r}\sin(\theta_{\rm d}/2)]^2\}$. The function $f(\theta_{\rm d})$ is known from weak-scattering theory and, e.g., describes interference in Rayleigh--Gans scattering~\cite{EAkkermans_1st_ed2007}. For clouds larger than the wavelength, the forward lobe is the dominant contribution to the elastic scattering for detection angles smaller than $\theta_{\rm c}=2/[\ln(N)k_0\sigma_{\rm r}]$, {\em i.e.}, close to the exact forward direction. The forwardly directed emission can be seen in Fig.~\ref{fig:cloud}(b) where $\ob{S_{\rm el}(\theta_{\rm d})}/N^2$ is plotted for different $N$ and fixed $b_0$ for $s\approx2$. The magnitude of the forward emission is given by~\cite{footnote:EPAPS}
\begin{align}\label{eq:Stheta0}
\ob{S_{\rm el}(\theta_{\rm d}=0)}/S_0=\frac{\pi N^2}{1+s}\left[\ob{n}^{(0)}+(1-s)\ob{n}^{(1)}\right]\delta(\delta\omega),
\end{align}
consisting of a non-interacting part and the first-order correction. Interestingly, Eq.~\eqref{eq:Stheta0} shows how the first-order correction to the forward-scattering lobe can be expressed in terms of the corresponding first-order correction to the steady-state population $n^{(1)}$ of Eq.~\eqref{dipole_blockade}. Surprisingly, the elastically scattered intensity, which is proportional to the frequency integral of the elastic spectrum, is not proportional to the atomic population as is otherwise found in the linear optics~\cite{bienaime_fsc12a} and single atom~\cite{LMandel_1995} limits. This signifies that detection of scattered light is not a direct measure of the atomic population.  

For $s\approx 2$ and $b_{\De}\approx0.004$ as used in Fig.~\ref{fig:cloud}(b) the correction to the elastic spectrum due to the dipole-dipole interaction is on the order of $10^{-5}$. While this correction appears to be small for the elastic spectrum, we will see that in the inelastic spectrum the cooperative effects are considerable.

\section{Inelastic spectrum} We now turn to the inelastic component of the spectrum of Eq.~\eqref{eqn:Spec}, $\ob{S_{\rm in}}$, and study how the Mollow triplet is affected by interatomic interactions. While some limits of the steady-state population and elastic spectrum can be investigated in the linear-optics regime, the inelastic spectrum is a truly nonlinear quantum optical phenomenon that calls for the theory reported in this Letter.

\begin{figure}[t!]
	\centering
		\includegraphics[width=0.45\textwidth]{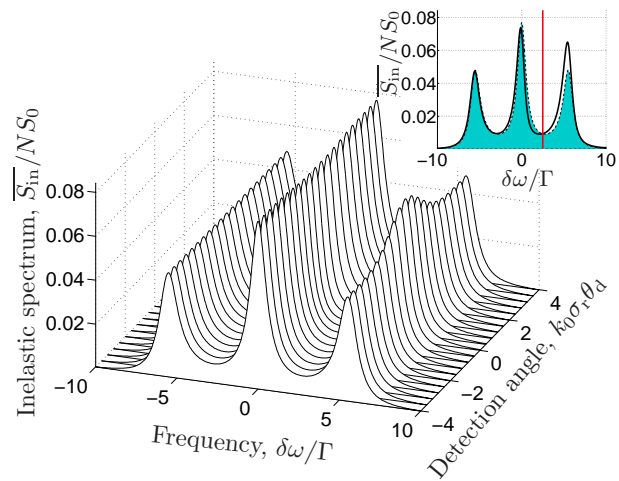}
	\caption{(Color online) Inelastic fluorescence spectrum versus rescaled detection angle, $k_0\sigma_{\rm r}\theta_{\rm d}$, with $N=25000$, $k_0\sigma_{\rm r}=5000$, $\OR/\Gamma=5$, and $\Delta/\Gamma=-2.5$. Inset: comparison of the asymmetric inelastic fluorescence spectrum for $\theta_{\rm d}=0$ (black solid line) with the symmetric non-interacting atoms spectrum (black dashed line with green shaded area). The vertical red line shows $\omega_a$.}
	\label{fig:NAtom_S_in_angle_w_inset}
\end{figure}

For non-interacting atoms an angle-independent inelastic emission pattern is found, which is simply $N$ times the single-atom Mollow triplet. Interestingly, when interactions are included also the inelastic spectrum becomes angle dependent. This is shown in Fig.~\ref{fig:NAtom_S_in_angle_w_inset}, where the inelastic fluorescence spectra of a cloud of $N=25000$ atoms and size $k_0\sigma_{\rm r}=5000$ for strong ($\OR/\Gamma=5$), off-resonant ($\De/\Gamma=-2.5$) driving are depicted for several detection angles. While the number of atoms $N=25000$ is by far too large for usual numerical calculation methods, it is relevant for experimental settings. The cloud size corresponds to the experimental value of Ref.~\cite{bienaime_prl10a}, where a detuning ranging from $-1.9\Gamma$ to $-4.2\Gamma$ was used consistent with our $\De=-2.5\Gamma$.  

All spectra shown in Fig.~\ref{fig:NAtom_S_in_angle_w_inset} exhibit the typical three-peak structure of the single-atom Mollow spectrum, but also a spectral asymmetry, which is strongest in the forward direction, see inset. In contrast,  the single-atom Mollow spectrum is symmetric, even for off-resonant driving. The observed asymmetry depends on the laser-atom detuning and gives an increase of the sideband peak closest to the bare atomic transition frequency $\omega_{\rm a}$. The $\sim35\%$ enhancement of the peak at the bare atomic frequency, as shown in Fig.~\ref{fig:NAtom_S_in_angle_w_inset}, is a result of cooperative effects showing the importance of including the dipole-dipole interactions when dealing with the fluorescence spectrum. 

We can now appreciate the pronounced cooperative features in the inelastic spectrum as compared to the elastic spectrum: for non-interacting atoms the elastic spectrum already shows an $N$-times enhanced forward-directed peak, but the inelastic spectrum does not. While the elastic peak is slightly modified due to interatomic interactions for the inelastic spectrum the same interactions create a forwardly-directed peak. In more detail, the dipole-dipole interactions create interatomic correlations such that the forward-directed interference pattern is built up in the inelastic spectrum.

\section{Conclusions, discussion, outlook} In conclusion, we have shown that dipole-dipole interactions in clouds of cold atoms affect their optical properties in the strong-driving regime, even for dilute clouds. We found analytical corrections to the steady-state population and to the fluorescence spectrum under strong driving. The analysis allows connecting the cooperative decay rate and Lamb shift with the Green function governing photon propagation. We found that, while a spherical Gaussian distributed cloud exhibits decreased atomic excitation, also a cooperatively increased atomic excitation is possible. Moreover, we have shown that cooperative scattering persist in the Mollow triplet, which is a hallmark of non-classical scattering of light by two-level systems. The cooperative effect gives rise to an angle-dependent spectrum and is most pronounced in the forward direction where it manifests itself as an enhancement of the sideband nearest to the atomic-transition frequency.

While we have considered the simplest model for the atoms, the scalar two-level model, it is worth noting that the approach used in this work can be generalized, e.g., to account for the full vectorial nature of the atom-light scattering and the near-field components of the dipole-dipole coupling. We are confident that our results for strongly driven dilute clouds will stimulate the study of denser clouds where interatomic interactions are expected to be even more important.

\section{Acknowledgments} We acknowledge K.~M{\o}lmer for valuable discussions. J.~R.~O. acknowledges financial support from the Otto M{\o}nsted foundation during his stay at the Institut Non-Lin{\'e}aire de Nice.

%

\end{document}